\documentclass[conference]{IEEEtran}
\IEEEoverridecommandlockouts
\usepackage{cite}
\usepackage{amsmath,amssymb,amsfonts}
\usepackage{algorithmic}
\usepackage{graphicx}
\usepackage{textcomp}
\usepackage{xcolor}
\usepackage{gensymb}
\usepackage{caption} 
\usepackage{newtxtext,newtxmath}
\def\BibTeX{{\rm B\kern-.05em{\sc i\kern-.025em b}\kern-.08em
    T\kern-.1667em\lower.7ex\hbox{E}\kern-.125emX}}
\begin{document}

\title{Thwarting Cybersecurity Attacks with Explainable Concept Drift
}

\author{\IEEEauthorblockN{Ibrahim Shaer}
\IEEEauthorblockA{\textit{Department Electrical and Computer Engineering} \\
\textit{University of Western Ontario}\\
London, Canada \\
ishaer@uwo.ca}
\and
\IEEEauthorblockN{Abdallah Shami}
\IEEEauthorblockA{\textit{Department of Electrical and Computer Engineering} \\
\textit{University of Western Ontario}\\
London, Canada \\
abdallah.shami@uwo.ca}
}

\maketitle

\begin{abstract}
Cyber-security attacks pose a significant threat to the operation of autonomous systems. Particularly impacted are the Heating, Ventilation, and Air Conditioning (HVAC) systems in smart buildings, which depend on data gathered by sensors and Machine Learning (ML) models using the captured data. As such, attacks that alter the readings of these sensors can severely affect the HVAC system operations impacting residents' comfort and energy reduction goals. Such attacks may induce changes in the online data distribution being fed to the ML models, violating the fundamental assumption of similarity in training and testing data distribution. This leads to a degradation in model prediction accuracy due to a phenomenon known as Concept Drift (CD) — the alteration in the relationship between input features and the target variable. Addressing CD requires identifying the source of drift to apply targeted mitigation strategies, a process termed drift explanation. This paper proposes a Feature Drift Explanation (FDE) module to identify the drifting features. FDE utilizes an Auto-encoder (AE) that reconstructs the activation of the first layer of the regression Deep Learning (DL) model and finds their latent representations. When a drift is detected, each feature of the drifting data is replaced by its representative counterpart from the training data. The Minkowski distance is then used to measure the divergence between the altered drifting data and the original training data. The results show that FDE successfully identifies 85.77 \% of drifting features and showcases its utility in the DL adaptation method under the CD phenomenon. As a result, the FDE method is an effective strategy for identifying drifting features towards thwarting cyber-security attacks. 

\end{abstract}

\begin{IEEEkeywords}
Concept Drift, Data Drift, HVAC systems, Explainable Concept Drift, Drift Localization. 
\end{IEEEkeywords}
\section{Introduction}
The development of autonomous systems, such as smart buildings and autonomous vehicles, has accelerated due to the enhancements in information and communication technologies (ICT) and increased storage capabilities of sensors \cite{SHAER2023100882}. These advancements contributed to the widespread deployment of Internet of Things (IoT) devices, which form the foundation for Machine Learning (ML) models crucial for the functionality of smart buildings. \cite{SHAER2023100882}. However, the integrity of the data used by ML models is at risk due to cyber-security attacks. These attacks can inject new data points or tamper with the sensory readings, resulting in data distributions that deviate from the training data distributions. This condition violates the data distribution similarity assumption for ML models, resulting in the degradation of their performance. The change in the data distribution is referred to as a Covariate Shift (CS), which can affect the mappings between the features and response variable, which is referred to as the Concept Drift (CD) \cite{mallick2022matchmaker}. 

In the context of smart buildings, CD manifests dire consequences to the decision-making process of Heating, Ventilation, and Air Conditioning (HVAC) systems. IoT devices play a crucial role in providing a digital footprint of the spatial environment, which is essential for the effective operation of HVAC systems. However, when these sensors are compromised during a cyber-security attack, they will reflect an inaccurate state of the environment. As a result, the HVAC system will be oblivious to the true state of the environment, and will, consequently, operate to the detriment of the building operator's objectives, such as reducing energy consumption and ensuring occupant comfort \cite{SHAER2023100882}.

The mitigation of the CD effect resulting from the cyber-security attack is instrumental in the restoration of the proper operation of the HVAC systems. In addressing CD, the traditional ML cycle of training and testing is extended to include drift detection and adaptation. Identifying the source of the drift is essential to implement effective mitigation strategies to thwart these attacks. Finding the drift source is referred to as \textit{drift localization}. For example, should a particular set of sensor readings be identified as the main source of the drift, a decision might be made to either ignore these readings or cut their Internet access to prevent any unnecessary HVAC system activation.  

Few prior works explored the explainability in the CD context. The first attempt by Demsar and Bosnic \cite{demvsar2018detecting} used off-the-shelf eXplainable Artificial Intelligence methods (XAI) while Hinder \textit{el al.} \cite{hinder2023model} included a more extensive experimental setup with XAI. In the same context, the works of \cite{yang2021cade, hinder2022contrasting, hinder2022localization, mallick2022matchmaker} identify the features mostly affected by drifts that gave rise to the field of \textit{drift localization}. The main issue of these methods is their experimentation with tree-based methods with inherent explainability capability, while the ones investigating Deep Learning (DL) methods are constrained with classification models. To address these shortcomings, we propose a novel approach named Feature Drift Explanation (FDE), which is applied to a regression task using a DL model. The key contributions of this paper are as follows:
\begin{itemize}
    \item Propose the first method for explaining CD in DL regression tasks within smart building environments, termed FDE;
    \item Introduce a novel CD explainability method that leverages the reconstruction capability of Autoencoders and their latent representations;
    \item Develop a distance metric to assess the closeness of drift and normal activation values of the regression DL model;
    \item Validate the effectiveness of FDE using a real-world dataset, demonstrating its capability to identify drifting features with an accuracy exceeding 85\%. 
\end{itemize}

The rest of the paper is organized as follows. Section II presents the related work and its limitations. Section III details the proposed framework. Section IV explains the experimental setup. Section V analyzes the obtained results. Section VI concludes the paper. 

\section{Related Work}
This work intersects the topics of the CD of DL models in smart building environments and the explainability of CD phenomena. As such, the literature related to these subjects is surveyed to draw comparisons and highlight the novelty of our proposed solution.

Several works addressed the CD phenomena in smart building settings, mainly connected to load and energy prediction \cite{mariano2022analysis, jagait2021load, bayram2023lstm, ji2021enhancing} and anomaly detection \cite{fenza2019drift}. However, our work differs in two main aspects. First, the proposed CD methodologies are applied to a univariate regression model, while our method considers a multi-variate use case. Second, none of these works integrated explainability, due to its univariate nature. This extends the explainability issue from identifying the drifting timestamp, which is part of the drift detection module, to identifying the drifting features. 

Few works investigated the explainability of the CD phenomena manifested through the emergence of the drift localization literature. In that regard, Hinder \textit{et al.} \cite{hinder2022contrasting, hinder2022localization, hinder2023model} are all works that significantly advanced this field. Counterfactuals representing post-drift examples close enough to pre-drift examples with a different label were utilized in \cite{hinder2022contrasting} to explain the drifting features. Similarly, \textit{MatchMarker} \cite{mallick2022matchmaker} finds the training batch which is the closest to the drifting sample for inference. Its extension \cite{hinder2023model} expands on the explainability of CD by integrating a corpus of available XAI methods, synonymous with the work by Desmar and Bosnic \cite{demvsar2018detecting}. Dimensionality reduction and clustering techniques were utilized in \cite{manias2021concept} to identify drifting nodes. CADE \cite{yang2021cade} is also a pioneering work that contributes to drift detection and its explanation via the application of contrastive learning as a basis to identify a drift and find the set of features affected by the drift. Similarly, Adams \textit{et al.} \cite{adams2023explainable} attempts to explain CD in process mining. The two main shortcomings of these works are as follows: these works have pre-dominantly applied their methodology on tree-based methods that are inherently explainable and when applied to a DL model none of the works considered regression tasks.  

\begin{figure}[ht]
    \centering
    \includegraphics[scale=0.37]{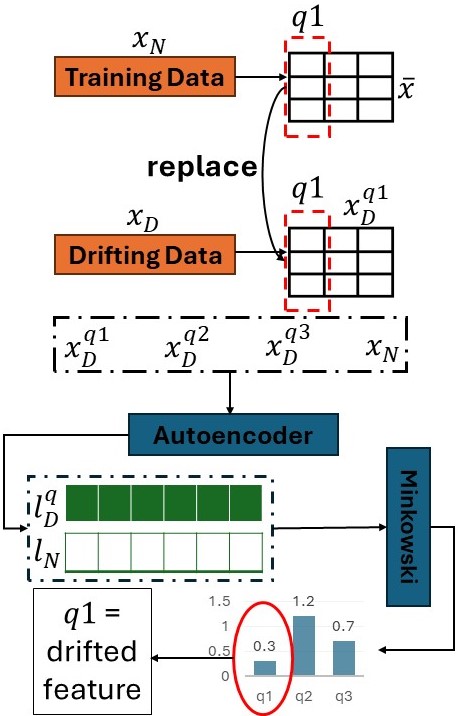}
    \caption{FDE Methodology}
    \label{fig:fde_methodology}
\end{figure}

\section{Methodology}
The pipeline for mitigating the CD phenomena requires three main steps: drift detection, explanation, and adaptation. We focus on the drift explanation aspect by defining the Feature Drift Explanation (FDE) module, depicted in Figure \ref{fig:fde_methodology}, that identifies the set of features affected by the drift. The design of FDE considers its potential future integration into a drift adaptation module for DL models. These adaptation strategies include fine-tuning or embedding techniques \cite{pasquadibisceglie2023darwin}. 

A deviation in the performance of a DL model or changes in data distribution triggers the drift mitigation strategies. To experiment with a DL model on time-series (TS) data, we created a 1-dimensional Convolutional Neural Network (1D-CNN) model that experiences a degradation in performance during a cyber-security attack. After the drift occurrence, a drift detection module raises alerts that activate the subsequent modules in the mitigation strategy pipeline. Data distribution-based drift detection methods fit the requirements of the covariate shift; however, they provide poor explainability which is important for the drift adaptation strategies of DL models \cite{lu2018learning}. Therefore, we resorted to performance-based detection methods, such as ADWIN \cite{bifet2007learning, Elena-TriLS-2022}, that raise drift alerts upon performance degradation.  

Before explaining the FDE module, it is important to highlight its main two goals: \textbf{(1)} find the features that are affected by the cyber-security attack, and \textbf{(2)} pinpoint the drifting DL model weights to aid the fine-tuning process of the DL adaptation method. The different steps that are part of the FDE module are designed to align with these defined goals. After the completion of the 1D-CNN model training, the first step in the FDE process collects the activation of the first convolutional layer ($a$) of this model when inferring the training dataset. Here, these activations provide a rich source of information about the lagged relationship between the set of features and the inter-feature relationships constructed by the 1D-CNN model to predict the response variable. 

The intuitive next step is to leverage a distance metric that measures the difference between the activations under normal ($a_N$) and drifting ($a_{D}$) behaviour. However, these distance metrics perform poorly with high dimensional data and induced sparsity. Additionally, the contributions of each feature can be reflected differently in $a$, which means that one-to-one distance computation can misrepresent the underlying relationships and variations. Therefore, it is imperative to utilize compression techniques to infer the latent representation ($l_a$), with a dimension $p$, that preserves important information about $a$, which solidifies the credibility of the computed distance metric. Principal Component Analysis (PCA) and Autoencoders (AE) are prime candidates to execute this compression. However, the non-linearity imposed by the CNN model violates the linear relationship assumption of PCA, favouring the compression procedure of AE \cite{shaer2023corrfl}. The AE, parameterized by $\theta$, is trained using the $a_N$ of each input $x \in R^{Q\times1}$, such that $Q$ represents the dimensions of the sequence data, with a loss function that minimizes the reconstruction error between the $a_N$ and the reconstructed activations referred to as $\Hat{a_N}$. Implementing AE in the FDE pipeline enables the inference of anomalous $a_D$ after the drift occurrence, facilitating the drift adaptation process. The loss function is as follows: 
\begin{equation}
    \min_{\theta}  \mathop{{}\mathbb{E}}_{a} \left\lVert a_N - \Hat{a_N} \right\rVert^{2}_{2}
\end{equation}

The final step is finding the set of features that were mainly affected by the cyber-security attack. Inspired by \cite{yang2021cade, mallick2022matchmaker}, this process is applied by comparing the drifting data points with the training data points. However, compared to these approaches, we introduce two main changes. First, the latent representation leveraged in CADE \cite{yang2021cade} is a disposable module when considering the drift adaptation process for DL models. 
Second, funding the closest raw data points in classification is archaic to a regression, due to shifts in decision boundaries. Therefore, in our use case, we first assume that the data points with the standardization mean and variance are used as representatives of the training dataset ($\Bar{x}$). $\Bar{x^{q}}$ represents the values of feature $q$ for $\Bar{x}$. The latent representations of all the training data are retained in a set $L_N$, such that $l_N$ is the latent representation of $a_N$. During a drift, we replace each feature $q$ at a time for a drifting point $x_D$ with its respective value from $\Bar{x}$, resulting in $x_{D}^{q}$. After that, the latent representation $l_{D}^{q}$ is obtained.  The Minkowski distance $M_q$ is computed between each $l_{D}^{q}$ and $l_N$, resulting in multiple values of $M_q$, each corresponding to a feature $q$. The feature with the smallest average distance to all training data is the drifting feature. This stems from the fact that replacing the drifting features with the training representative features brings them closer to the latent space of the training data. $M_q$ is computed as follows such that $l_{D}^{qp}$ represents the latent representation of instances $a_D$, such that feature $q$ is replaced with $x_{q}$ at dimension $p$,
is as follows: 
\begin{equation}
    M_q = (\sum_{i=1}^{p}(\lVert l^{p}_N - l_{D}^{qp} \rVert))^{\frac{1}{p}}
\end{equation}

\section{Experimental Setup}
This section explains the dataset used to validate the proposed approach, the parameters of FDE, and the drift simulation setup. 

\subsection{Dataset Details}
The dataset used for this study gathered five environmental features for over a year in an office environment in the Nordic climate of Finland using six sensors\cite{rasanen_2020_4311286}. This dataset was captured in 13 rooms belonging to two spatial settings: a conference room and a cubicle. We confine our analysis to a conference room that fits 12 people, referred to as room00, as a proof-of-concept for FDE. The deployed sensors capture the CO\textsubscript{2} concentration (ppm), pressure (hPa), temperature (\degree C), humidity (\%), and passive infrared count (PIR). Since the PIR count reflects the movement levels, its value was calculated by aggregating these levels in five-second intervals. 

This experimental procedure follows the pre-processing steps applied by the original authors \cite{kallio2021forecasting}. First, the values of different sensors of room00 were aggregated and aligned based on their timestamps. After that, the data was resampled to a one-minute interval and median interpolation was applied for co-occurring values from different sensors. Lastly, chunks of 60-minute continuous data intervals were retained for forecasting purposes. These continuous sections were assigned specific frame values to maintain the integrity of predictions and the data sequence pre-processing. The dataset was split into training and testing sets, whereby lagged values of environmental features with a 5-minute history window, while the response variable is the future 5-minute CO\textsubscript{2} concentration levels. The testing set consists of more challenging prediction instances comprising CO\textsubscript{2} level changes that are more significant than the $80^{th}$ quantile of the overall change \cite{kallio2021forecasting}. 

\subsection{Neural Network Models' Description}
In our methodology, two neural network structures are utilized. The first structure is the 1D-CNN model, which is an appropriate choice for DL models in the context of explainability. The training dataset was processed to follow a sequence structure that enables the 1D-CNN model to infer the lagged relationship between a set of features. The resulting dataset is of the form (number of timesteps, number of features), such that the number of timesteps represents the history time window. 

The tuning of kernel size is crucial for the performance of the 1D-CNN model \cite{tang2022omni}; however, we leave this procedure for eager practitioners as it has minimal effect on this paper's contribution. Therefore, to simplify the kernel size selection procedure, it is equal to 1 to account for each feature value in every timestep. The number of channels for the first convolutional layer determines the resultant activations', which is required for the next step in our analysis pipeline. We select 32 channels for this layer, resulting in 32 activation values per timestep. When flattened, this amasses 160 activation values for each data point. The AE is tasked with finding the best latent representation of activations. Toward that end, two main requirements were integrated into the design of the AE. The first requirement is that the number of input and output neurons should match the activations' dimensions for proper reconstruction. The best latent representation is attained with sufficient stages of compression, which emulates the 1D-CNN model structure. We consider a compression ratio of 90 \%, resulting in a latent dimension of 16.

\subsection{Drift Simulation and Evaluation Criteria}
 We assume that the data stream representing the testing dataset is split by the drift detection module into non-drifting and drifting sections. The proposed method is evaluated on the drifting sections to identify the set of features affected by the drift. To that end, we simulate a cyber-security attack that alters the values of one of the five environmental features by replacing their values with their outliers defined as 2 standard deviations ($\sigma$) away from their mean ($\mu$) value. This emulates a sudden CD that occurs during a cyber-security attack. The drift scenarios are summarized in Table \ref{tab:feature_stats}.

\begingroup
\setlength{\tabcolsep}{7pt} 
\renewcommand{\arraystretch}{1.5}
\begin{table}[]
\centering
\begin{tabular}{|c|c|c|}
\hline
\textbf{Feature}     & \pmb{$\mu$}   & \pmb
  {$\sigma$} \\ \hline
Temperature & 23.38  & 0.74               \\ \hline
Humidity    & 29.1   & 13.25              \\ \hline
Pressure    & 1006   & 12.12              \\ \hline
CO\textsubscript{2}         & 430.02 & 65.4               \\ \hline
PIR         & 0.31   & 1.46               \\ \hline
\end{tabular}
\caption{Feature Statistics}
\label{tab:feature_stats}
\end{table}
\endgroup

The alteration in the values of all and not a subset of environmental features is meant to showcase our method's ability to recognize all types of drifts. Since the 1D-CNN model is built using all the available features, it is expected the contribution of each lagged feature towards the prediction result is uneven. Therefore, the alteration of some features would not necessarily affect the prediction quality, eluding the performance-based drift detection modules. Despite this fact, the attack is still taking place, jeopardizing the smart buildings' environment. These conditions exhibit the possibility of different types of drifts, referred to as virtual and real drift \cite{oliveira2021tackling}. The real drift changes the relationship between the features and the response variable, whereas the virtual drift does not affect this relationship \cite{oliveira2021tackling}. 

We use the following evaluation criteria to validate our proposed method: (1): Mean Absolute Error (\textbf{MAE}) that evaluates the 1D-CNN model and is utilized by the drift detection module to raise drift alerts, (2): Mean Squared Error (\textbf{MSE}) the minimizes the reconstruction error between the input and output activations, and (3): average accuracy for correct identification of drift feature. The developed models and evaluation criteria were built using PyTorch \cite{paszke2019pytorch} python library. These models were evaluated on a Windows PC with a 3.00 GHz 24-core AMD Threadripper processor, 128 GB of RAM, and 16 GB Nvidia GeForce RTX 3060 Ti GPU. The code will be made available on a GitHub repository. 

\section{Results}
This section outlines the results of the experimental setup. It first shows the performance of the base models of FDE. After that, the contribution of the FDE in identifying the drifting features is detailed.  Lastly, the utility of the FDE module toward drift adaptation of DL models is presented. 

\subsection{Base FDE Model Evaluation}
While the evaluation of the base FDE module, constituting the 1D-CNN model and the AE model is not central to this paper's contribution, it is imperative to highlight their performance on the current problem. The 1D-CNN model is evaluated based on its predictive performance of CO\textsubscript{2} levels in the future 5 minutes using MAE and its inference time (ms/batch). Similarly, the AE model is evaluated based on its reconstruction error using MSE and its inference time (seconds/batch).

\begingroup
\setlength{\tabcolsep}{7pt} 
\renewcommand{\arraystretch}{1.5}
\begin{table}[]
\centering
\begin{tabular}{|c|c|c|}
\hline
\textbf{Model}  & \textbf{Performance} & \begin{tabular}[c]{@{}c@{}}\textbf{Inference Time} \\ \textbf{(ms)}\end{tabular} \\ \hline
1D-CNN & 2.4         & 0.017                                                             \\ \hline
AE     & 0.0006      & 0.037                                                              \\ \hline
\end{tabular}
\caption{FDE Models' Performance}
\label{tab:model_fde_per}
\end{table}
\endgroup

Table \ref{tab:model_fde_per} summarizes the evaluation of both models, averaged over 5 runs. Both models perform well in their respective domains. The low MAE of the 1D-CNN model shows this model successfully mapped the existing relation between the lagged environmental features and the future CO\textsubscript{2} concentration levels. Therefore, when a drift occurs, the performance degrades by disrupting the established relationship between the input and output features. Similarly, the low MSE for the AE model exhibits this model's utility in reconstructing the activations of the first convolutional layer. The low inference time also suggests that the models are suitable for real-time drift explanation capability. 

\subsection{FDE Assessment}

\begin{figure}[ht]
    \centering
    \includegraphics[scale=0.26]{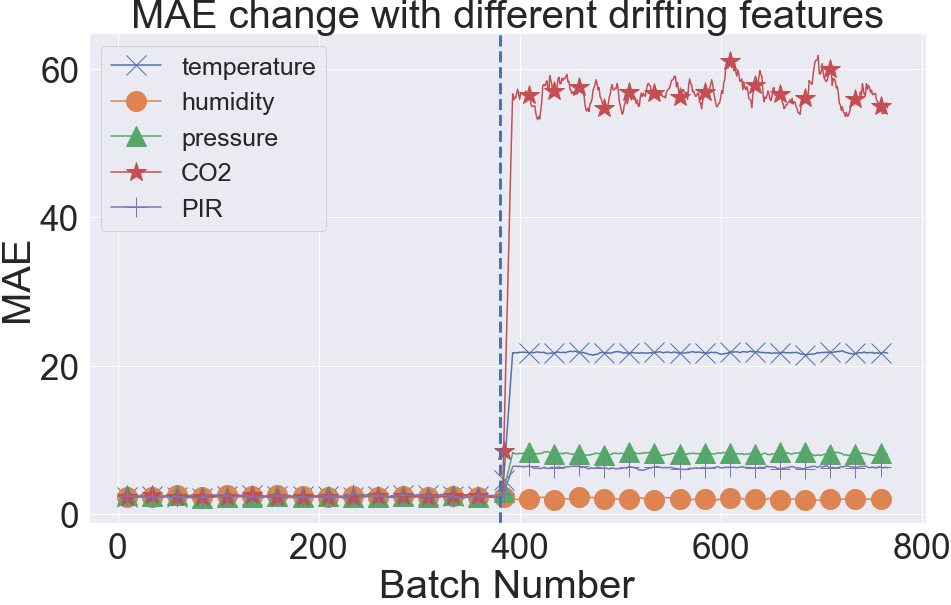}
    \caption{MAE for different drifting features}
    \label{fig:mae_drifting}
\end{figure}

Figure \ref{fig:mae_drifting} depicts the changes in MAE of the 1D-CNN models upon changing the values of the five environmental features, averaged over 30 runs, such that each data point represents the MAE of five batches of data. The horizontal line points to the time when the drift occurs. We observe drastic differences exist in MAE before versus after the drift, with prominent differences in the MAE based on the drifting feature. For example, only one feature has little to no effect on the MAE, upon simulating a drift for the humidity feature. This represents a virtual concept drift, as no effects are witnessed on the prediction performance. This type of drift can also extend to cases whereby the PIR and pressure features were undergoing a drift. The detection of a drift depends on the tolerable deviation of the MAE.

The alteration of the temperature and the CO\textsubscript{2} levels reflected in prominent changes in the MAE of the 1D-CNN model. For the temperature, the MAE increases more than 10-fold while for CO\textsubscript{2}, the increase is almost 30-fold compared to pre-drift MAE. This result shows that the contribution of each feature towards the predictions is not uniform, which necessitates methods that can monitor data distribution and performance changes. Espousing a hybrid approach is beneficial to detect different types of drift and infer the sensors which are under cyber-security attack. This approach would protect the HVAC systems from these attacks. 

\begingroup
\setlength{\tabcolsep}{7pt} 
\renewcommand{\arraystretch}{1.5}
\begin{table}[]
\centering
\begin{tabular}{|c|c|}
\hline
Feature     & \begin{tabular}[c]{@{}c@{}}Detection\\ Accuracy\end{tabular} \\ \hline
Temperature & 100 $\pm$ 0                                                     \\ \hline
Humidity    & 93.8 $\pm$ 0.024                                                \\ \hline
Pressure    & 100 $\pm$ 0                                                     \\ \hline
CO\textsubscript{2}         & 65.3 $\pm$ 0.06                                                 \\ \hline
PIR         & 69.75 $\pm$ 0.06                                                \\ \hline
\end{tabular}
\caption{Detection Accuracy of FDE}
\label{tab:fde_accuracy}
\end{table}
\endgroup

The next step is evaluating the accuracy of identifying the drifting features, which measures the distance between the latent representation of the altered drifting data point and those of normal points. Table \ref{tab:fde_accuracy} summarizes the detection accuracy results for all drifting features averaged over 30 experiments. Different observations can be inferred based on the drifting features. Good detection performances are achieved for the temperature, humidity, and pressure features. This suggests that FDE successfully detected drifting features and that the latent representation distance of the activations is representative of these drifting features. Additionally, it emphasizes FDE's virtual drift detection capability,  resulting from the drifting humidity feature and real drifts associated with the remaining feature. 

\begin{figure}[ht]
    \centering
    \includegraphics[scale=0.26]{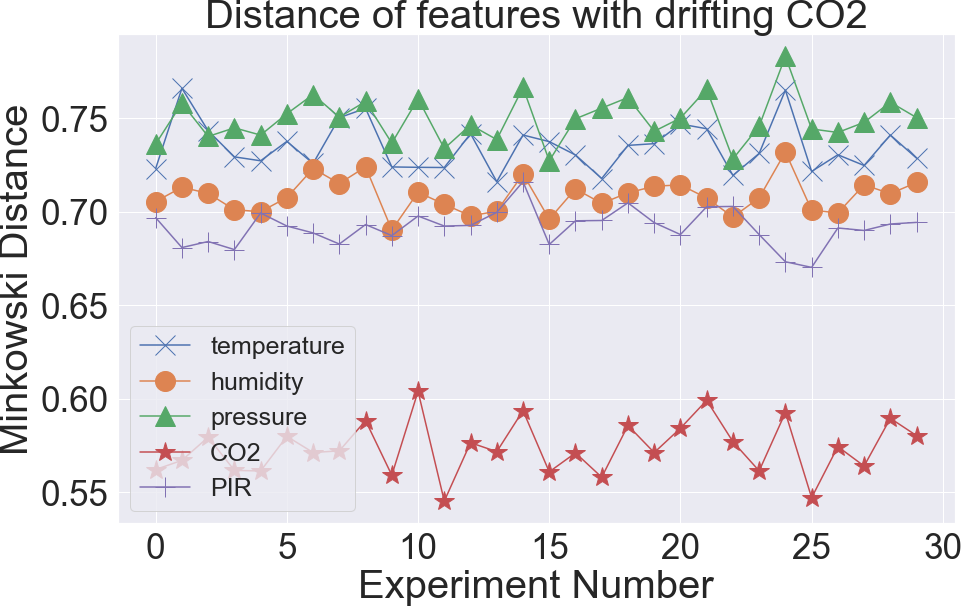}
    \caption{Minkowski distance of different features with drifting CO\textsubscript{2}}
    \label{fig:drifting_co2}
\end{figure}

The detection accuracy is less impressive for CO\textsubscript{2} and PIR count compared to the stellar performance of other features. These results can be attributed to many reasons. First, there is a high correlation between the CO\textsubscript{2} and PIR count \cite{shaer2023hierarchical}, which explains why both of these features exhibit the same results. Second, the detection accuracy depends on the latent's space ability to capture the relationships existing in the activations of the first convolutional layer. Since the lagged CO\textsubscript{2} features display the highest fidelity with the CO\textsubscript{2} level prediction as depicted in Figure \ref{fig:mae_drifting}, the latent representation is too reductive to represent the variations associated with the CO\textsubscript{2} activation changes. In other words, the compression ratio of the latent representation is high (90\%) to fully capture the activation information related to the CO\textsubscript{2} levels. 

The distances between the latent representations for each feature, when sensors collecting CO\textsubscript{2}s level are under attack, are depicted in Figure \ref{fig:drifting_co2} averaged over all the drifting samples. On average, the distances associated with CO\textsubscript{2} alteration are significantly lower than its counterparts, suggesting the success of the FDE approach. Additionally, in most experiments, the closest distance is associated with PIR, which aligns with our previous analysis. Therefore, to improve the detection results for CO\textsubscript{2} and PIR drifts, it is imperative to experiment with different compression ratios to strike the right balance between detection accuracy and the issues that arise with the high-dimensional space when calculating the Minkowski distance. 

\subsection{The utility of FDE for DL Model Adaptation}
\begin{figure}[ht]
    \centering
    \includegraphics[scale=0.26]{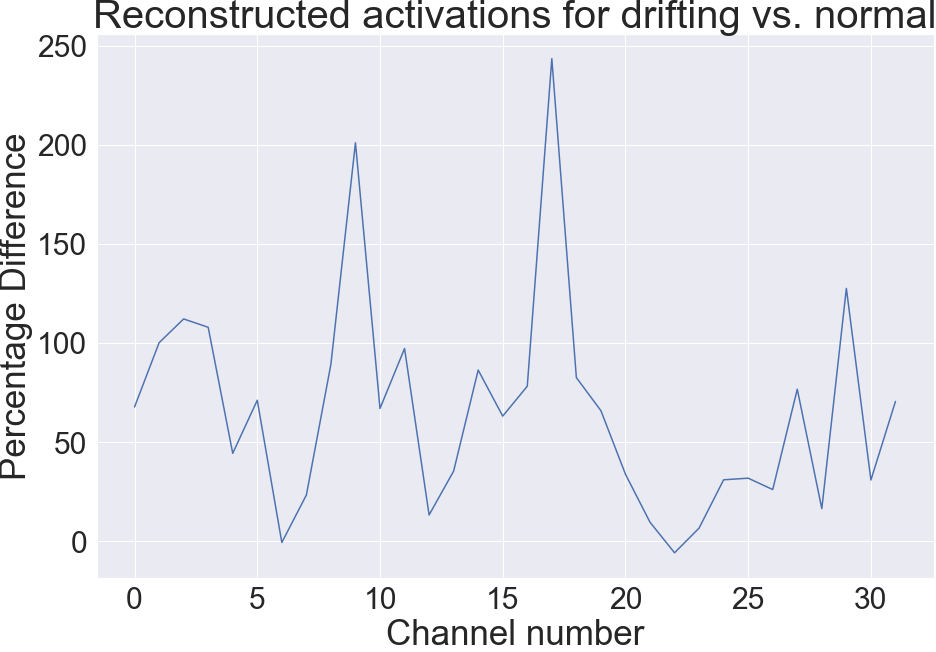}
    \caption{Percentage difference of reconstruction loss for drifting vs. normal activation instances}
    \label{fig:mse_reconstruction}
\end{figure}

The design of the FDE module considered its future integration as part of the DL model adaptation after encountering a drift. This subsection explains the insights FDE endows for any model adaptation methods, in terms of inferring which weights need to be fine-tuned to improve the model's performance. Figure \ref{fig:mse_reconstruction} displays the percentage difference of the reconstruction loss for drifting vs. normal activation instances. The drifting feature here is the CO\textsubscript{2} levels. To obtain these values, the reconstruction losses of the activations of the normal and drifting instances are computed via the AE model. After that, the percentage differences are calculated, taking the normal instances as a reference. Since the activations' dimension is 160 $(Channels \times Features)$, we aggregated the values to obtain the differences for the 32 channels for a compressed representation. 

Examining closely Figure \ref{fig:mse_reconstruction} shows that most of the drifting activations are associated with higher MSE compared to their normal counterparts. This demonstrates that AE can detect anomalous channels, which emphasizes its utility. Since the AE was not trained on drifting samples, it tries to find the closest examples it learned as a reconstructed activation for the drifting examples. This attempt increases the MSE of channels that are mostly affected by the drift. As such, the percentage differences highlight the channels that need to be fine-tuned to incorporate the new concept into the old one. Instead of re-tuning the whole model, the FDE pinpoints the channels that require calibration.

\section{Conclusion}
The widespread adoption of communication technologies has been pivotal in developing autonomous systems crucial to our well-being. However, these systems are vulnerable to cyber-security attacks. In smart buildings, HVAC systems rely extensively on sensors and ML models to automate their operations. Cyber-security attacks can tamper with sensor data, causing a concept drift - a phenomenon that degrades the performance of ML models' predictions. In this paper, we focus on enhancing the explainability of concept drift, allowing human operators to identify and address vulnerabilities effectively. Accordingly, we propose a new method named Feature Drift Explanation (FDE), which isolates the affected features, facilitating the attack mitigation strategies. FDE utilizes Autoencoders to find latent representations of feature activations to measure the distances between normal and drifting features. FDE successfully identifies the drifting features, displaying high detection accuracy. In addition, the insights gathered from the FDE operation can serve as a rich landscape to facilitate the DL adaptation methods. For future work, we will investigate different compression techniques for the AE method, applying this method to other datasets, and incorporating concept drift adaptation strategies. 

\bibliographystyle{IEEEtran}
\bibliography{refs}
\end{document}